\documentclass[epsf,twocolumn,showpacs,preprintnumbers]{revtex4}
\usepackage{graphics}
\usepackage{graphicx}
\usepackage{dcolumn} 
\usepackage{bm}     
\usepackage{epsfig}
\begin{document}

\title{Born Effective Charges and Infrared Response of LiBC }
\author{Kwan-Woo Lee and W. E. Pickett}
\affiliation{Department of Physics, University of California, Davis, CA 95616}
\date{\today}
\begin{abstract}
Calculations of the zone center optical mode frequencies (including 
LO-TO splitting), Born effective charges Z$^*_{\alpha\alpha}$ for each
atom, dielectric constants $\epsilon_{0}$ and $\epsilon_{\infty}$, 
and the dielectric response in the infrared, using density functional
linear response theory, are reported.  Calculated Raman modes  
are in excellent agreement with experimental values (170 cm$^{-1}$ and 
1170 cm$^{-1}$), while it will require better experimental data to 
clarify the infrared active mode frequencies.  The Born effective
charges Z$^*_{\alpha \alpha}$  
(i) have surprisingly different values for B and C, and
(ii) show considerable anisotropy.
Relationships between the effective charges and LO-TO splitting are
discussed, and the predicted reflectivity in
the range 0 -- 1400 cm$^{-1}$ is presented.  These results hold
possible implications for
Li removal in LiBC, and C substition for B in MgB$_2$.
\end{abstract}
\maketitle

\section{Introduction}
Discovery of superconductivity near 40 K in MgB$_2$,\cite{akimitsu}
and the rapid development 
in the understanding of the microscopic mechanism (a specific type 
of electron-phonon coupling) has spurred extensive work 
on its properties and the discovery of closely related materials.
Interest in LiBC and its Li-deficient derivatives has been spurred 
by calculational evidence\cite{rosner1} that the electron-phonon coupling will be
stronger, and consequently the superconducting transition temperature
is expected to be higher, than MgB$_2$. 
LiBC is isostructural to, and isovalent with, MgB$_2$ to the extent 
possible in a ternary compound, as we discuss  more below.

LiBC was first reported by W\"orle {\it et al.},\cite{worle} who 
provided the crystal structure, and also reported hole doping by removal 
of Li that changed
it from reddish in color to black, and increased the conductivity.  
Recently there
has been renewed interest in synthesizing this compound and studying its
properties.  Bharathi {\it et al.} reported synthesis for varying starting
concentrations of Li,\cite{bharathi} but did not report any samples 
becoming metallic. 
Hlinka and coworkers reported Raman scattering measurements
on few-micron size crystallites,\cite{hlinka} 
finding modes at 170 cm$^{-1}$ and
1176 cm$^{-1}$, which were recently confirmed by Renker {\it et al.}
(171 and 1167 cm$^{-1}$).\cite{renker}  
Souptel {\it et al.} reported synthesis of samples for 
several flux concentrations, and like Bharathi {\it et al.} found
conductivities characteristic of lightly doped semiconductors.\cite{souptel}
Beginning optical studies on LiBC have been reported by Pronin {\it et al.}
using an infrared microscope on small crystals,\cite{pronin}
providing infrared (IR) reflectivity in the 400-1900 cm$^{-1}$ range and
measurements of the dielectric behavior at lower frequencies.
We will return below to discussion of the implications of the Raman and
IR data for the vibrational properties of LiBC.

As for possible doping (de-intercalation of Li), all of these reports 
are probably consistent with very low doping, too low to drive the material
conducting (in spite of the broad bands that promote doped-hole
conduction).  The susceptibility data indicate small concentrations of
local moments, typical of semiconductors with point defects.  
Zhao, Klavins, and Liu, in their efforts at Li de-intercalation,
observed evaporation of Li upon vacuum annealing, with no sign of 
superconductivity in the annealed samples.\cite{zhao} 
The amount of Li remaining in the sample has not yet 
been characterized, however (xrays are not sensitive enough), so 
definitive results remain to be obtained.

The purpose of this paper is to obtain in more detail the character of
the intrinsic (insulating) phase of LiBC.  Phonon frequencies calculated
by linear response techniques using the full-potential linear muffin-tin
orbital (FP-LMTO) method were reported by An {\it et al.}\cite{jan2}
In this paper we use related by extended methods to calculate not only
the phonon frequencies (again, with different codes) but the dynamic
effective charge tensor Z$^*_{\alpha}$, the longitudinal -- transverse
(LO-TO) mode splittings, oscillator strengths of the IR active phonons,
and related dynamic dielectric behavior.  Since Li is known, from 
calculation, to be ionic in this compound, one could anticipate
Z$^*_{Li} \approx +1$, and because 
B and C lie side-by-side in the periodic table, that Z$^*_B \approx$
Z$^*_C$ and of course Z$^*_B$ + Z$^*_C =$ -Z$^*_{Li}$ from the acoustic
sum rule.  We find much more interestng behavior, however; Z$^*_B$ and
Z$^*_C$ are 
of opposite sign, and large -- they behave like strongly charged cation
and anion, respectively.  We analyze this behavior, and also comment on
the possible relevance this behavior has for the difficulty in
de-intercalating Li from LiBC, and also for C doping in MgB$_{2-x}$C$_x$.

\section{Calculational Methods}

The calculations were carried using the ABINIT code with Troullier-Martins 
pseudopotentials\cite{1,2}, Teter parametrization\cite{3} 
of the Ceperley-Alder exchange-correlation potential, 
and 216 $\mathbf{k}$-points in the irreducible wedge of the Brillouin zone.  
The kinetic energy cutoff for the plane waves was 60 Hartree.  
The $\it{f}$- sum rule (relative to unity) \cite{8} was satisfied 
with to 1.0005, which is one measure that the basis set is 
effectively complete.
The experiment lattice constants $(a=2.752 { }\AA,{ }c=7.058 { }\AA )$ 
\cite{worle} of LiBC (space group $P6_{3}/mmc$, No. 194) were used.
The calculation of effective charges follows the formalism of Gonze
and Lee\cite{gonze} as implemented in the ABINIT code.

\begin{figure}[bt]
\psfig{figure=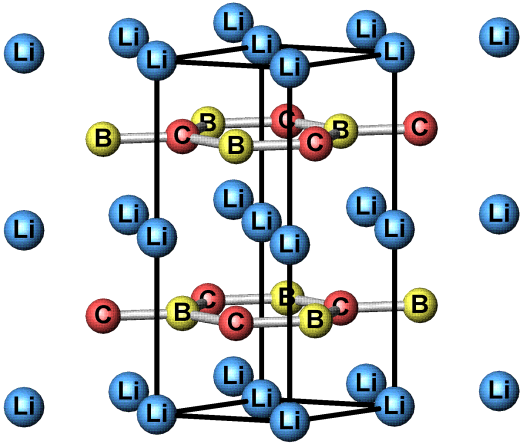,width=7.0cm,angle=-0}
\caption{\label{structure} 
Crystal structure of LiBC, showing the alternating B-C arrangement in the
graphene layers.  B and C atoms also alternate along the $\hat c$ axis,
resulting in two layers and two formula units per primitive cell.  The
Li ions reside in the interstitial positions between B-C hexagonal
rings above and below.}
\end{figure}

The crystal structure of LiBC, pictured in Fig. \ref{structure}, is a
direct generalization of that of MgB$_2$.  The B and C atoms form a flat
graphene sheet, with B and C atoms alternating around each hexagonal
unit.  The stacking of B-C layers is alternating, such that each B has
C neighbors along the $\hat c$ direction, and similarly for C.  This
stacking, which doubles the unit cell volume over that of MgB$_2$, 
indicates that B-C bonding along $\hat c$ (to the extent that it occurs)
is preferable to B-B and C-C bonding.  Possibly this stacking results
in a favorable Madelung energy, although real static charges associated
with B and C separately are almost impossible to define or calculate.   
The Li ions lie in the interstitial site (center of inversion)
between the centers of the 
hexagonal units above and below, coordinated at equal distances with six B
atoms and six C atoms.

\section{Results}
\subsection{Vibrational Frequencies}

The character of the various zone center
modes in LiBC were presented earlier by 
An {\it et al.}\cite{jan2}  For completeness, we repeat them here, 
with our calculated frequencies (in cm$^{-1}$), with macroscopic 
electric field contributions neglected; those corrections and 
longitudinal optic (LO) - transverse optic (TO)  splittings are 
discussed below.
Factor group analysis\cite{factor} at the $\Gamma$ point yields 15 
optical modes : 
$2A_{2u} + 2B_{1g} + B_{2u}$ with motions along the $\hat c$ direction,
and $2E_{1u} + 2E_{2g} + E_{2u}$ with motions in the $a-b$ plane
(the latter type are all two-fold degenerate $E$ modes).
The $2A_{2u}$
and $2E_{1u}$ are infrared active and the $2E_{2g}$ modes are Raman
active.\cite{hlinka,pronin,renker,jan2}

\begin{itemize}
\item $\omega$=169: E$_{2g}$, B-C layers sliding against each
  other 
\item $\omega$=299: B$_{2g}$, B-C layers beating against each other
  along $\hat c$ 
\item $\omega$=292: E$_{2u}$, Li layers sliding against each other 
\item $\omega$=346: E$_{1u}$, Li layers sliding against the B-C 
   layers 
\item $\omega$=407: A$_{2u}$, Li layers beating against the B-C layers,
  along $\hat c$ 
\item $\omega$=510: B$_{1u}$, Li layers beating against each other
  along $\hat c$
\item $\omega$=803: A$_{2u}$, B-C puckering mode, all B atoms
  move oppositely
  to all C atoms, Li sites become inequivalent
\item $\omega$=829: B$_{2g}$, B-C puckering mode, B moves with C atoms
  above/below it, Li sites remain equivalent
\item $\omega$=1143: E$_{1u}$, B-C bond stretching mode, the two
  layers are out-of-phase
\item $\omega$=1153: E$_{2g}$, B-C bond stretching mode, layers in phase
\end{itemize}

Our calculated values are also provided in Table \ref{table1}, where
LO-TO splittings, where they arise, are also reported.
Compared with the observed Raman active modes 
at 170 and 1176 cm$^{-1}$\cite{hlinka} and 171 and 1167 
cm$^{-1}$,\cite{renker} 
the calculated values, 169 and 1153 cm$^{-1}$, are in excellent agreement.

\begin{table}[bt]
\caption{Zone center optical modes of LiBC. The first column
($\mathbf{E}=0$) is for no electric field,
the second is for the field lying in the plane
($\mathbf{E}\perp \hat{c}$), and the
third is for $\mathbf{E} \parallel \hat{c}$.
The $2A_{2u}$
and $2E_{1u}$ are infrared active and the $2E_{2g}$ modes are Raman active.
$\Delta {\omega^2}$ = $\omega_{LO} ^2$ - $\omega_{TO}^2$ }
\begin{center}
\begin{tabular}{|l||l|l|l||l|lr|} \hline\hline
 modes & \multicolumn{3}{c|}{Phonon frequency ($cm^{-1}$)} &
                          $\sqrt{\Delta {\omega^2}}$ \\ \cline{2-4}
       & $\mathbf{E}$ =0 &  $\mathbf{E} \parallel a-b$ &
         $\mathbf{E} \parallel \hat{c}$ & ($ cm^{-1}$)  \\ \hline\hline
 $E_{2g}$  &{ }169 &~ { }169 &{ }169 &  \\ \cline{2-4}
           &{ }169 &~ { }169 &{ }169 &  \\ \cline{1-4}
 $E_{2u}$  &{ }292 &~ { }292 &{ }292 &  \\ \cline{2-4}
           &{ }292 &~ { }292 &{ }292 &  \\ \cline{1-4}
 $B_{2g}$  &{ }299 &~ { }299 &{ }299 &  \\ \cline{1-4}
 $E_{1u}$  &{ }346 &~ { }346 &{ }346 &  \\ \cline{2-5}
           &{ }346 &~ { }$\bf{371}$ &{ }346  &~~ 135   \\ \hline
 $A_{2u}$  &{ }407 &~ { }407 &{ }$\bf{507}$ & ~~  302   \\ \hline
 $B_{2u}$  &{ }510 &~ { }510 &{ }510 &  \\ \hline
 $A_{2u}$  &{ }803 &~ { }803 &{ }$\bf{828}$ & ~~ 200    \\ \hline
 $B_{1g}$  &{ }829 &~ { }829 &{ }829 &  \\ \cline{1-4}
 $E_{1u}$  & 1143 & ~ 1143 & 1143 &  \\ \cline{2-5}
           & 1143 & ~ $\bf{1235}$ & 1143 &   ~~ 469   \\ \hline
 $E_{2g}$  & 1153 & ~ 1153 & 1153 &  \\ \cline{2-4}
           & 1153 & ~ 1153 & 1153 &  \\ \hline\hline
\end{tabular}
\end{center}
\label{table1}
\end{table}

The LO-TO splitting for displacement parallel to the layers occurs
for the ``layer sliding'' mode
at 346 cm$^{-1}$ ($\Delta \omega = 25$ cm$^{-1}$), and for the B-C
``bond stretching'' mode  
at 1143 cm$^{-1}$ ($\Delta \omega = 92$ cm$^{-1}$).  
For $\hat c$ polarization,  
the ``layer beating''
mode is at 407 cm$^{-1}$ ($\Delta \omega = 100$ cm$^{-1}$), and the B-C 
``puckering'' mode at 803 cm$^{-1}$ is split by 
only $\Delta \omega = 25$ cm$^{-1}$.  The large LO-TO splittings of the 
``layer beating'' and ``bond stretching'' modes suggest they involve
large effective charges, as we confirm below.

Data on IR active modes have been published by Pronin {\it et al.}
\cite{pronin} who fit reflectivity data with modes at 540, 620, 700 (less
clear) and 1180 cm$^{-1}$.  The highest (and strong in the data) mode is
close to our calculated value (TO 1143 cm$^{-1}$, LO 1235 cm$^{-1}$),
but the others are difficult to assign.  The data, with three or four 
positions of structure, apparently include contributions from $\hat c$ axis
polarization, $a-b$ plane polarization, and all angles between.
Bharathi {\it et al} have presented IR absorption spectra\cite{bharathi}
with main
peaks at $\sim$380, 950, and 1200 cm$^{-1}$.  The lower and upper of these
are not far from our calculated values for E$_{1g}$ modes 
(Table \ref{table1}), but
the nearest IR mode to the 950 cm$^{-1}$ peak is our A$_{2u}$ mode at
803-828 cm$^{-1}$.

\subsection{Effective Charges and Oscillator Strengths}
In a non-cubic lattice the Born (dynamical) 
effective charge of an ion becomes a tensor.  
In hexagonal symmetry such as in LiBC,
the tensor is diagonal and reduces to two values 
Z$^*_{xx} = $ Z$^*_{yy} \equiv$ Z$^*_{\parallel}$ and
Z$^*_{zz} \equiv$ Z$^*_{\perp}$.  These charges, which show considerable 
anisotropy in LiBC, are given 
in Table II.
The acoustic sum rule $\sum_{s} Z_{s}^* =0 $ is fulfilled to within 0.01,
suggesting well converged calculations.
In spite of the fact that they are neighbors
in the periodic table and that both B and C are 
comfortable forming the graphene structure layer, when they alternate
in this graphene layer their effective charges are vastly different: 
substantial in size but different in sign.
Given their similarities, the effective charge sum rule, and that Li is
ionized, one might perhaps expect values close to +1, 0, and -1 
(or possibly +1, -$\frac{1}{2}$, and -$\frac{1}{2}$)for Li, B, and 
C, respectively.  They are, on (angular) average, much closer to the values
+1, +2, and -3, which are not so far from the bizarre 
ionic configuration (only one with closed shells) of
+1, +3, and -4 for Li, B, and C respectively that would reflect closed
shells on all ions.  It must be recognized, of course, that the Z$^*$
values are dynamic only, and reflect the effects of 
covalency with respect to some reference
ionic value; however, the reference values here are unclear (except for
Li, which clearly should be +1).

Due to the expected strong covalency in the B-C bonds, it perhaps is not
surprising that the B and C effective charges are strongly altered from
(smaller) reference ionic values.  For $\hat c$ axis polarization,
the Li value itself is unusual (almost 50\% larger than its nominal
value), indicating Li is definitely involved in the interlayer coupling.
We return to this important feature below.
This partial Li covalency is probably connected with the difficulty
in deintercalating it from the LiBC lattice.\cite{souptel,renker}

Another view of the charge response can be obtained from the ``mode
effective charges'' defined by Gonze and Lee\cite{gonze}, which for the
symmetry of LiBC can be written for each polarization as 
\begin{eqnarray}
\vec Z^* = \sum_{\kappa}  \frac{Z^*_{\kappa} \vec U_{\kappa}}
  {\sum_{\kappa'} \vec U^*_{\kappa'} \cdot \vec U_{\kappa'}}.
\end{eqnarray}
Here each ion effective charge is weighted by the mode eigenvector 
that is normalized without the ion mass factor that occurs in the usual
normalization of $U_{\kappa\alpha}$ ($\kappa$ is the atom index, $\alpha$
is the Cartesian coordinate).  The mode effective charge vectors 
indicate how strongly a mode will couple to an electric field, and are
listed in Table III for the IR active modes in LiBC.  For 
the $\hat c$ axis polarized modes, $|\vec Z^*|$ is 3.00 for the (softer)
layer beating mode and 2.10 for the (harder) B-C puckering mode.  
For planar displacements, the (soft) layer sliding mode has the value
1.52, and the (hard) B-C bond stretching mode value is 5.50.  These trends
follow those found by Zhong {\it et al.} in perovskites,\cite{zhong} where 
displacements that modulated the ``covalent'' bonding produced the
largest mode effective charge.  The value of 3.00 for the layer beating
mode reflects the substantial Li ``covalent'' character for $\hat c$ axis
displacements.

\subsection{Infrared Response}
The static electronic dielectric constants are calculated to be
$\epsilon_{\infty}^{\perp}=12.95$, $\epsilon_{\infty}^{\parallel}=11.24$.
These values tend to be overestimated in LDA calculations such as those
used here, presumably due to the underestimation of the bandgap.
Dielectric constants are also somewhat dependent (at the $\sim$5\% level)
on the choice of pseudopotential.
According to the generalized Lyddane-Sachs-Teller (LST) relation, 
\begin{eqnarray}
\epsilon_{0} = \epsilon_{\infty} \prod_m 
      \frac{\omega^2_{LO},m}{\omega^2_{TO,m}}
\label{genLST}
\end{eqnarray}
which is used separately for each polarization,
the static dielectric constants
$\epsilon_{0}$ of both directions are $17.4$ and $18.5$, respectively.
In the GHz range ($\sim$0.03 cm$^{-1}$), $\epsilon \approx 35 $ was
reported by Pronin {\it et al},\cite{pronin} but this value is not
necessarily expected to be intrinsic.

\begin{table}[!bt]
\begin{center}
\caption{The Born effective charges of LiBC. The acoustic sum rule
  $\sum_{s} Z_{s}^* =0 $ is fulfilled to within 0.01.  $\overline{Z^*}$
  = (1/3) Tr$\mathbf{Z^*}$}
\begin{tabular}{|l||llr|}    \hline \hline
                   & $Z^{*}(Li) $ &$Z^{*}(B)$  & $Z^{*}(C)$ \\ \hline \hline
$\mathbf{Z^*_{\parallel}}$ &~ 0.81    &~2.37     & -3.17~  \\
$\mathbf{Z^*_{\perp}}$     &~ 1.46    &~0.61     & -2.07~  \\ \hline
$\overline{Z^*}$           &~ 1.03    &~1.78     & -2.80~  \\ \hline \hline
\end{tabular}
\end{center}
\label{table2}
\end{table}

The relationship between the LO-TO splitting and the Born effective charges 
is given by 
the relation\cite{gonze,resta} (which holds for parallel and perpendicular
vibrations separately),
\begin{eqnarray}
\sum_{m} [ \omega_{LO,m}^2  -  \omega_{TO,m}^2] & = &
\frac{4\pi}{\epsilon_{\alpha \alpha}^{\infty}  V_{\circ}}
 \sum_{\kappa} {({Z_{\kappa, \alpha\alpha}^* e)^2}\over 
   {M_{\kappa}}} \nonumber \\
  &=& \sum_{\kappa} \Omega^2_{ion,\kappa}.
\end{eqnarray}
In this relation, $m$ goes over the IR active modes of the given polarization, 
$M_{\kappa}$
the ionic mass of aton $\kappa$, $V_{\circ}$ the volume of the primitive
unit cell, and the right hand side has been expressed in terms of screened
ionic plasma frequencies $\Omega_{ion,\kappa}$ that indicate how much each 
ion contributes to the combined strength of the resonances.  For in-plane
polarization, Li, B, and C contribute 6\%, 26\%, and 58\%; the Li ion
has minor effect and the response is dominated by B-C bond stretching.  
For $\hat c$ axis polarization, the contributions are
43\%, 5\% and 51\% respectively; here the B is almost irrelevant, and the
Li contribution is comparable to that of C.  

\begin{table}
\begin{center}
\caption{Mode effective charges of the IR active optical modes}
\begin{tabular}{|l||llr|} \hline \hline
~~~~  modes       &  \multicolumn{3}{c|}{Directions}  \\ \cline{2-4}
                  & ~ a   & ~ b  &  c~~              \\ \hline \hline
layer sliding $E_{1u}$ &{ }0.65 &-1.37    &{ }0.00          \\
                   &{ }1.37 &{ }0.65  &{ }0.00        \\ \hline
layer beating $A_{2u}$ &{ }0.00 &{ }0.00  &{ }3.00        \\ \hline
B-C puckering $A_{2u}$ &{ }0.00 &{ }0.00  &-2.10          \\ \hline
bond stretching $E_{2u}$ &-5.50 &-0.10      &{ }0.00             \\
                    &-0.10 &{ }5.50    &{ }0.00   \\ \hline \hline
\end{tabular}
\end{center}
\label{table3}
\end{table}

The contributions can be examined mode by mode.  
The dielectric function can be expressed in terms of contributions 
from the IR active modes ($m$) as
\begin{eqnarray}
\frac{\epsilon(\omega)}{\epsilon^{\infty}} & = &
 1 + \frac{4\pi}{\epsilon^{\infty} V_{\circ}} \sum_m \frac{S_m}
     {\omega_{TO,m}^2 - \omega^2} \nonumber \\
 &=&1 +  \sum_m \frac
     {\omega_{LO,m}^2 - \omega_{TO,m}^2}
     {\omega_{TO,m}^2 - \omega^2}.
\label{epsomega}
\end{eqnarray}
The second line follows from the relation between the mode oscillator
strength $S_m$ and the corresponding LO-TO splitting, provided in
Table \ref{table1}
in terms of $\Delta \omega^2$.
For in-layer polarization,
large LO-TO splitting of the bond stretching $E_{1u}$ 
mode at 1143 (1235) cm$^{-1}$
is the result of the large B and C effective charges; the splitting  
of the layer sliding 
mode at 346 (371) cm$^{-1}$ is an order of magnitude less.
For $c$ axis polarization, the effective charges are ``derived'' 70\%
from the 407 (507) cm$^{-1}$ layer beating mode with large LO-TO splitting 
involving the larger Z$^*$ for Li, 
and only 30\% derives from the 803 (828) cm$^{-1}$ 
B-C puckering mode. 

\begin{figure}[bt]
\psfig{figure=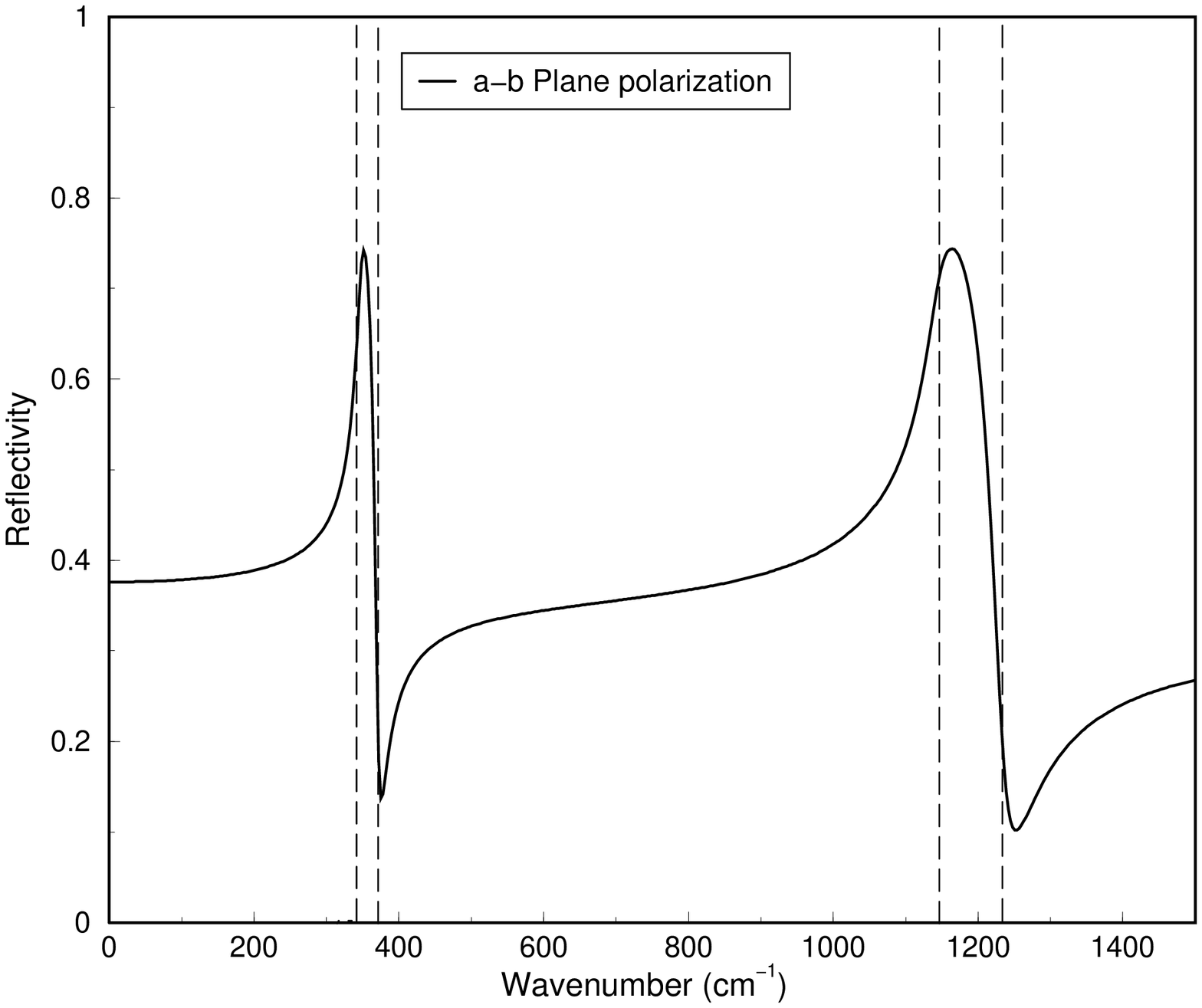,width=5.5cm,angle=-90}
\psfig{figure=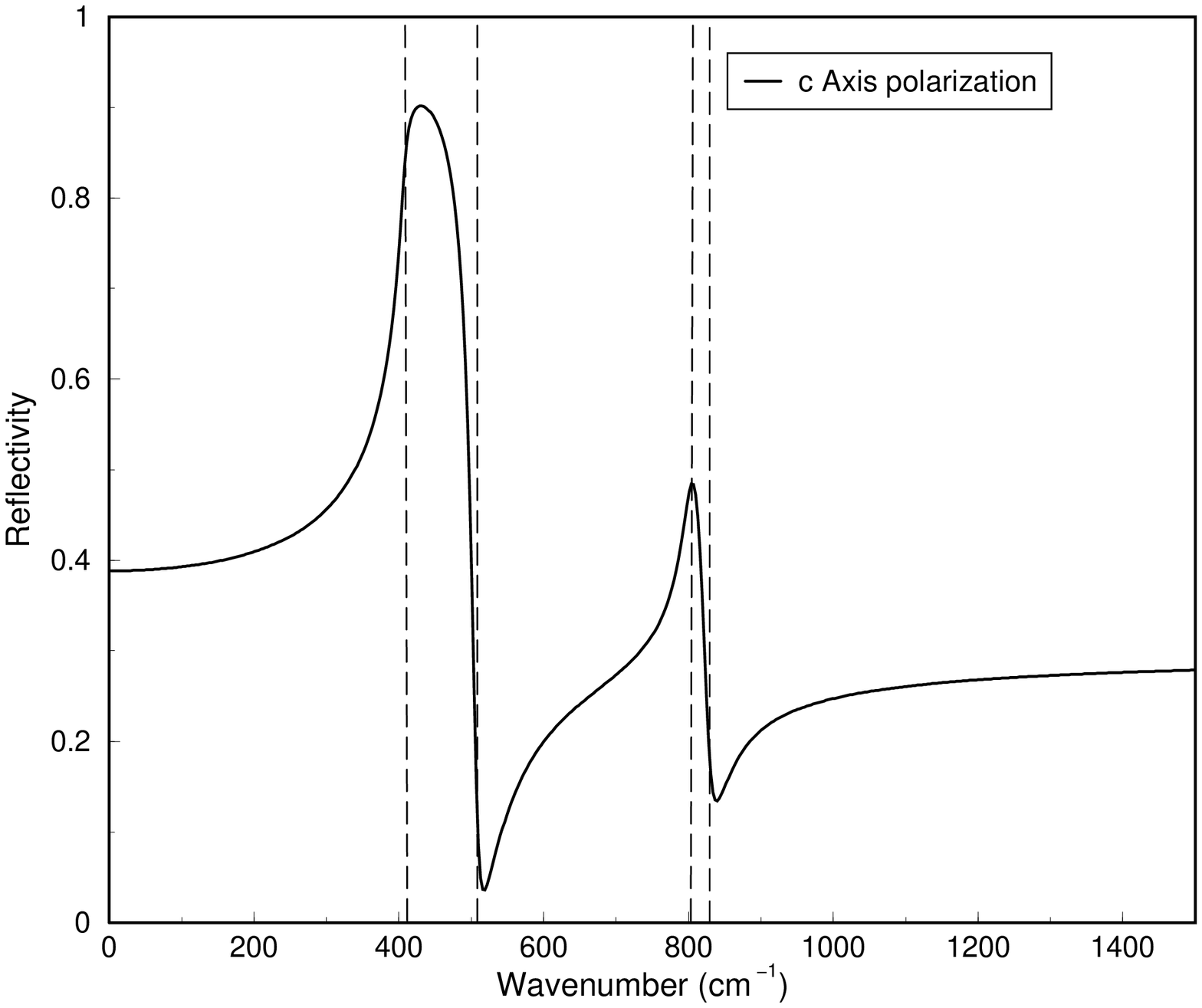,width=5.5cm,angle=-90}
\caption{\label{reflectivity}
Calculated reflectivity spectrum R$(\omega$) for LiBC for polarization
in the $a-b$ layer (top panel) and along the $\hat c$ axis (bottom
panel), calculated from Eq. \ref{epsomega2}.  
The dashed lines mark the positions of the TO and the LO modes
for each IR active mode (two of them for each polarization).  The
damping was chosen to be 3\% of the frequency for each mode.
}
\end{figure}

From Eq. \ref{epsomega} the zero frequency limit is given by
\begin{eqnarray}
\frac{\epsilon(\omega)}{\epsilon^{\infty}} -1 & = &
  \sum_m \Bigl(\frac {\omega_{LO,m}^2}{\omega_{TO,m}^2}-1 \Bigr)
\end{eqnarray}
which is different from the approximate ``generalized Lyddane-Sachs-Teller''
relation of Eq. \ref{genLST} that is often quoted.  It is not as different 
as it looks.  The ``textbook'' form of the
dielectric function in terms of its zeroes and poles is
\begin{eqnarray}
\frac{\epsilon(\omega)}{\epsilon^{\infty}} = \prod_m
  \frac{\omega_{LO,m}^2 -\omega^2}{\omega_{TO,m}^2 -\omega^2}.
\label{epsomega2}
\end{eqnarray}
While this expression has the same poles (at $\omega_{TO,m}$) as
Eq. \ref{epsomega}, the positions of its zeroes, at $\omega_{LO,m}$
in this expression, are different from the zeroes in the general expression
Eq. \ref{epsomega}.  However, zeroes have to occur between each of the
poles, and the zeroes of Eq. \ref{epsomega} are determined not only
by $\omega_{LO,m}$ but also by the polarization arising from other LO
modes with $m^{\prime} \neq m$.  The differences are small, however.
For in-plane polarization, the zeroes of $\epsilon(\omega)$ from
Eq. \ref{epsomega} are 367 cm$^{-1}$ and 1236 cm$^{-1}$ compared to the
calculate ``LO'' frequencies of 371 and 1235 cm$^{-1}$, respectively. 
(Note that the lower is pushed down, while the higher is increased, for
the case of two LO modes as we have here.)  
For $\hat c$ axis polarization, Eq. 
\ref{epsomega} gives 499 and 833 cm$^{-1}$ compared to the calculated
values of 507 and 828 cm$^{-1}$.  These differences will be difficult
to resolve experimentally.

\section{Discussion and Summary}
The calculated Raman-active frequencies are in excellent agreement with
the data of two groups.\cite{hlinka,renker}  The IR modes remain to 
be verified and understood.  The spectra are polarization dependent, 
and the spectra presented by Pronin {\it et al.}\cite{pronin} and
by Bharathi {\it et al.}\cite{bharathi} are not representative of pure
polarization data.  

Since it will be useful for comparison with single crystal data,
in Fig. 2 we present our predicted reflectivity $R(\omega)$ spectrum
for both in-plane and perpendicular polarization.  The damping (which
for a clean undoped sample
would be due to anharmonicity) has been chosen for each mode to be 3\%
of the frequency (zero damping curves are not very representative).
If the damping is not too large, and there is no reason to expect large
anharmonicity in LiBC, then both TO and LO frequencies can be obtained
with small uncertainty.

It was mentioned in the Introduction that calculations have indicated
that, if partial Li removal can be achieved, Li$_{1-x}$BC
should be a very good superconductor.  Such Li extraction is common in
many materials, being the process that forms the basis for Li batteries.
What our studies here have shown is that (primarily for $\hat c$ axis
displacements) Li shows considerable covalency with the B-C layer,
with Li-C coupling being the prominent feature.
This partial Li covalency is probably connected with the difficulty
that several groups have found\cite{souptel,renker} 
in de-intercalating it from the LiBC lattice.  In addition,
Cava, Zandbergen, and Inumaru reported briefly that 
LiBC seems to be highly resistant to
chemical doping.\cite{cava} 

The finding here that C is very different chemically from B in this 
system also carries some implications for C replacement of B in MgB$_2$.
Recent studies confirm that about 10\% of B can be replaced with C
while retaining the structure, and the superconducting transition temperature
remains high (T$_c$=22 K).\cite{canfield,taken,jorgensen}  
If C alloying in the B sublattice
were rigid band like, the additional 0.1 electron per cell would nearly 
fill the important $\sigma$ bands\cite{rosner} that are responsible for
superconductivity, and T$_c$ would be expected to vanish (or nearly so).
The great difference between C and B behavior in LiBC reflects strong
differences in bonding, and suggests something very different from
rigid band ``doping.''  Preliminary calculations for MgB$_{2-x}$C$_x$
(ordered supercells and virtual crystal)\cite{deepa}
indeed indicate substantial 
non-rigid-band behavior: C is really different from ``B with an extra
electron.''

Although calculations such as those described here have been found in
several systems to be accurate, it is highly desirable to obtain polarized
single crystal IR data to confirm our predictions.

\section{Acknowledgments}
W.E.P. acknowledges informative communicaitons with A. Loidl, K. Liu, and
J. Hlinka.  We have benefitted greatly from the ABINIT project, and
from personal communication with X. Gonze.  
This work was supported by NSF Grant No. DMR-0114818.

\newpage


\begin{thebibliography}{10}

\bibitem{akimitsu}J. Nagamitsu {\it et al.}, Nature {\bf 410}, 63 (2001).

\bibitem{rosner1}H. Rosner, A. Kitaigorodsky, and W. E. Pickett,
  Phys. Rev. Lett. {\bf 88}, 127001 (2002).

\bibitem{worle}M. W\"orle, R. Nesper, G. Mair, M. Schwarz, and H. G. von
  Schnering, Z. Anorg. Allg. Chem. {\bf 621}, 1153 (1995).

\bibitem{bharathi}A. Bharathi, S. J. Balaselvi, M. Premila, T. N. Sairam,
  G. L. N. Reddy, C. S. Sundar, and Y. Hariharan, Solid State Commun.
  {\bf 124}, 423 (2002).

\bibitem{hlinka}J. Hlinka, I. Gregora, J. Pokorny, A. V. Pronin, and A.
  Loidl, Phys. Rev. B {\bf 67}, 020504 (2003).

\bibitem{renker}B. Renker, H. Schober, P. Adelmann, P. Schweiss, K.-P.
  Bohnen, and R. Heid, cond-mat/0302036.

\bibitem{souptel}D. Souptel, Z. Hossain, G. Behr, W. L\"oser, and
  C. Geibel, Solid State Commun. {\bf 125}, 17 (2003).

\bibitem{pronin}A. V. Pronin, K. Pucher, P. Lunkenheimer, A. Krimmel,
 and A. Loidl, cond-mat/0207299.

\bibitem{zhao}L. Zhao, P. Klavins, and K. Liu, J. Appl. Phys. (2003, in press).

\bibitem{jan2}J. M. An, H. Rosner, S. Y. Savrasov, and W. E. Pickett,
  Physica B (2003, in press) [cond-mat/0209256].

\bibitem{1} N. Troullier and J. L. Martins, Phys. Rev. B {\bf 43}, 1993, (1991).

\bibitem{2} See URL 
 $http://www.abinit.org/ABINIT/Psps/$-- \
~~$LDA_TM/lda.html$

\bibitem{3} See the Appendix of S. Goedecker, M. Teter, and J. Hutter,
  Phys. Rev. B {\bf 54}, 1703 (1996).

\bibitem{8}Z. H. Levine and D. C. Allan, Phys. Rev. Lett. 
   {\bf 63}, 1719 (1989).

\bibitem{gonze} X. Gonze and C. Lee, Phys. Rev. B {\bf 55},
   10355 (1997).

\bibitem{factor}D.L Rosseau, R.P. Bauman and S.P.S Porto, J. Raman Spectroscopy
        {\bf 10}, 253, (1981).

\bibitem{zhong}W. Zhong, R. D. King-Smith, and D. Vanderbilt, Phys. Rev.
  Lett. {\bf 72}, 3618 (1994). Note that these authors used a somewhat
  different definition of the mode effective charge than was given by
  Gonze and Lee.\cite{gonze}

\bibitem{resta} R. Resta, M. Posternak, and A. Baldereschi, Phys. Rev. Lett.
            {\bf 70}, 1010 (1993).

\bibitem{cava}R. J. Cava, H. W. Zandbergen, and K. Inumaru, Physica C
  {\bf 385}, 8 (2003). 

\bibitem{canfield}R. A. Ribeiro, S. L. Bud'ko, C. Petrovic, and P. C.
  Canfield, Physica C {\bf 385}, 16 (2003).

\bibitem{taken}T. Takenobu, T. Ito, D. H. Chi, K. Prassides, and Y. Iwasa,
  Phys. Rev. B {\bf 64}, 134513 (2001).

\bibitem{jorgensen}M. Avdeev {\it et al.}, cond-mat/0301025.

\bibitem{rosner}H. Rosner, J. M. An, W. E. Pickett, and S.L. Drschsler,
  Phys. Rev. B {\bf 66}, 024521 (2002).

\bibitem{deepa}D. Kasinathan and W. E. Pickett, unpublished.

\end{thebibliography}
\end{document}